\documentclass[12pt]{iopart}
\pdfoutput=1
\usepackage{amssymb}
\usepackage{setstack}
\usepackage{palatino}
\usepackage{graphicx}
\usepackage{cite}
\def\dblone{\hbox{$1\hskip -1.2pt\vrule depth 0pt height 1.6ex width 0.7pt
                  \vrule depth 0pt height 0.3pt width 0.12em$}}
\begin{document}
\title{Equilibration of quantum hard rods in one dimension}
\author{S. Ji , C. Ates, J. P. Garrahan and I. Lesanovsky}
\address{School of Physics and Astronomy, The University of Nottingham, Nottingham, NG7 2RD, United Kingdom}
\ead{ppxsj1@nottingham.ac.uk,
igor.lesanovsky@nottingham.ac.uk}

\begin{abstract}
We study the out-of-equilibrium evolution of a strongly interacting quantum spin chain which is mapped on a system of  hard rods that are coherently deposited on and removed from a lattice. We show that this closed quantum system approaches an equilibrium steady state which strongly resembles a microcanonical ensemble of classical hard rods. Starting from the fully coherent evolution equation we derive a Master equation for the evolution of the number of hard rods on the lattice. This equation does not only capture properties of the equilibrium state but also describes the dynamical non-equilibrium evolution into it for the majority of initial conditions. We analyze this in detail for hard rods of varying size.
\end{abstract}

\maketitle

\section{Introduction}
A central objective of non-equilibrium statistical mechanics is the description of the time evolution of thermodynamic properties of macroscopic systems that are prepared far from equilibrium. In principle one should think that it is always possible to derive an effective equation of motion, e.g. a Fokker-Planck equation, that governs the non-equilibrium behavior of thermodynamic quantities, from the underlying microscopics. In general this is, however, a formidable task and even more so in isolated systems where there is no apparent system-bath decomposition or an immediately manifest hierarchy in the coupling strengths between two or more subsystems. Nevertheless, these isolated systems often exhibit a generic relaxation dynamics, in the sense that for a subset of observables the non-equilibrium evolution leads to a steady state which is well-described by a thermodynamic ensemble.

For classical systems the underlying microscopic Hamilton equations of motion are non-linear in the dynamical variables. Generically, this leads to chaos such that the system ultimately explores the entire phase space. The occurrence of thermalization is then made plausible by invoking the Ergodic Hypothesis \cite{re:85}. In contrast, the microscopic equation describing the dynamics of a closed quantum system is a linear wave equation for a complex valued wavefunction and observables are represented by Hermitian operators. Due to these mathematical differences it is not obvious which equivalent arguments would lead to thermalization in this quantum context.

Given this, a first approach to study thermalization of quantum systems would involve a direct calculation of their microscopic dynamical evolution. However, determining the exact quantum dynamics of an interacting many-body system is a formidable task. The microscopic approach of solving Schr\"odinger's equation is usually restricted to very small systems, since the exponential growth of the Hilbert space with the number of particles also entails a growth in the mathematical complexity of the problem that quickly becomes unmanageable. Circumventing this problem requires sophisticated numerical tools and clever mathematical approaches. It is fair to say that our understanding of the equilibration and thermalization of closed quantum systems is still far less developed than for their classical counterparts.

Recently, there is an ever growing interest in addressing the question of when and how these systems relax towards an equilibrium state when prepared in a non-stationary initial state \cite{baal+:06,posh+:09,leol+:10,chki:10,biko+:10,gomu+:11,pode+:11,buda+:11,baci+:11,pose+:11,yu:11,sces:12,geho+:12,olgo+:12,atga+:12}. This development is, on the one hand, sparked by the availability of state of the art experiments - especially in the field of ultracold gases - where the investigated systems are extremely well isolated from the thermal environment and thus constitute almost ideal realizations of closed systems \cite{kiwe+:06,hole+:07,wibe+:10,trch+:12,grku+:12}. These experiments do not only constitute ideal test-beds for new theoretical ideas and concepts but they also point to the fact that there is growing interest for a quantum theory of the dynamics of closed systems to adequately interpret experimental observations made in these thermally isolated setups.

On the other hand, recent successes in the theory of equilibration and thermalization of closed quantum systems have fueled the interest in this topic. A very insightful idea was formulated with the Eigenstate Thermalization Hypothesis (ETH) that links thermalization to the spectral properties of the quantum Hamiltonian \cite{de:91,sr:94}. In essence, ETH states that in a thermalizing system any many-body eigenstate contains a thermal state, i.e., that the expectation value of an observable $\mathcal{O}$ calculated in an eigenstate with energy $\varepsilon$ coincides with the thermal average taken in the microcanonical ensemble at same energy. This conjecture has numerically been verified for a number of model systems \cite{ridu+:08,ecko:08,pahu:10}.

Despite its success the ETH does not provide an answer to how relaxation takes place and does not provide any information on the non-equilibrium evolution which eventually drives the system into a steady state. Here we take a closer look at the equilibration of a closed, interacting many-body quantum system in the time domain, expanding on a recent work \cite{atga+:12}. To this end, we consider a simple yet generic one-dimensional spin model, which belongs to the class of Ising models in a transverse field and can be mapped on a system of quantum hard rods that are coherently removed from/deposited on a lattice. For this system we show that it is possible to derive analytically a Master equation which captures the non-equilibrium dynamics of the number of hard rods. We undertake a comparison of this effective description with numerically exact simulations of the fully quantum problem and find excellent agreement. In this way, the results of this work also contribute to recent efforts to derive effective diffusion equations that describe the non-equilibrium dynamics in closed quantum systems \cite{atga+:12,tiva+:12p,ge:12pc}.

This paper is organized as follows: In  \sref{sec:hamiltonian} we introduce a spin Hamiltonian that represents a quantum version of a lattice gas of hard rods. Subsequently, in \sref{sec:config_network}, we introduce a specific graphical representation of the Hilbert space of the quantum model - the configuration network - to conveniently illustrate the dynamics of this system. Using the insights from a statistical analysis of the network, we present the derivation of a Master equation that describes the dynamics of the number distribution function of hard rods in \sref{sec:master_equation}. In particular, we show that the rate coefficients of this Master equations can be calculated analytically. Moreover, we numerically investigate the time evolution and steady state distribution of the density of hard rods. Following this analysis, we compare in \sref{sec:numerics} the predictions of the Master equation with  the results obtained by numerically solving the many-body Schr\"odinger equation for our model system and find excellent agreement. Conclusions and an outlook are provided in  \sref{sec:outlook}. Unless stated otherwise, we set $\hbar =1 $ in this work.

\section{The system} \label{sec:hamiltonian}
The quantum system in which we are interested here is a chain of $L$ equally spaced, interacting spin-$1/2$ particles at nearest neighbor distance $a$ in a transverse magnetic field of strength $\Omega$. Its Hamiltonian is given by
\begin{eqnarray}
H_{\mbox{\scriptsize spin}} = H_\Omega+H_V = \Omega \sum_{j=1}^L \sigma_j^x + V \sum_{j=1}^L \sum_{i=1}^\lambda n_j n_{j+i},
\label{eqn:spin_H}
\end{eqnarray}
where, $\sigma_j^x = \left(\left|\uparrow \right> \left< \downarrow \right| + \left|\downarrow  \right> \left< \uparrow \right| \right)_j$ is a Pauli spin matrix effectuating spin flips of the $j$-th spin and $n_j = \left( \left|\uparrow  \right> \left< \uparrow \right| \right)_j$ is a projector on the up-state (excited state) of the $j$-th spin. The spin-spin interaction of strength $V$ depends on the spin-state as well as on the separation of the spins:  two spins interact (i) if both of them are simultaneously in the spin-up state and (ii) if they are separated by a distance that does not exceed a critical distance $\lambda$, which we will refer to as \emph{blockade radius}. We use this terminology because in this work we focus specifically on the regime $V/\Omega \to \infty$. This means that it is energetically forbidden to have two or more up-spins located within the critical radius $r_{\mathrm{c}}=\lambda\,a$, i.e. an up-spin blocks the excitation of spins to the up-state in its vicinity. The model defined by the Hamiltonian (\ref{eqn:spin_H}) is in fact of practical relevance, as it has been shown to capture the quantum dynamics of strongly interacting, laser-driven Rydberg atoms trapped in an optical lattice \cite{viba+:11,scch+:12p}. For more details we refer the reader to references \cite{olmu+:10,le:11,atle:12,homu+:12p}.

\begin{figure}
\includegraphics[width=1\textwidth]{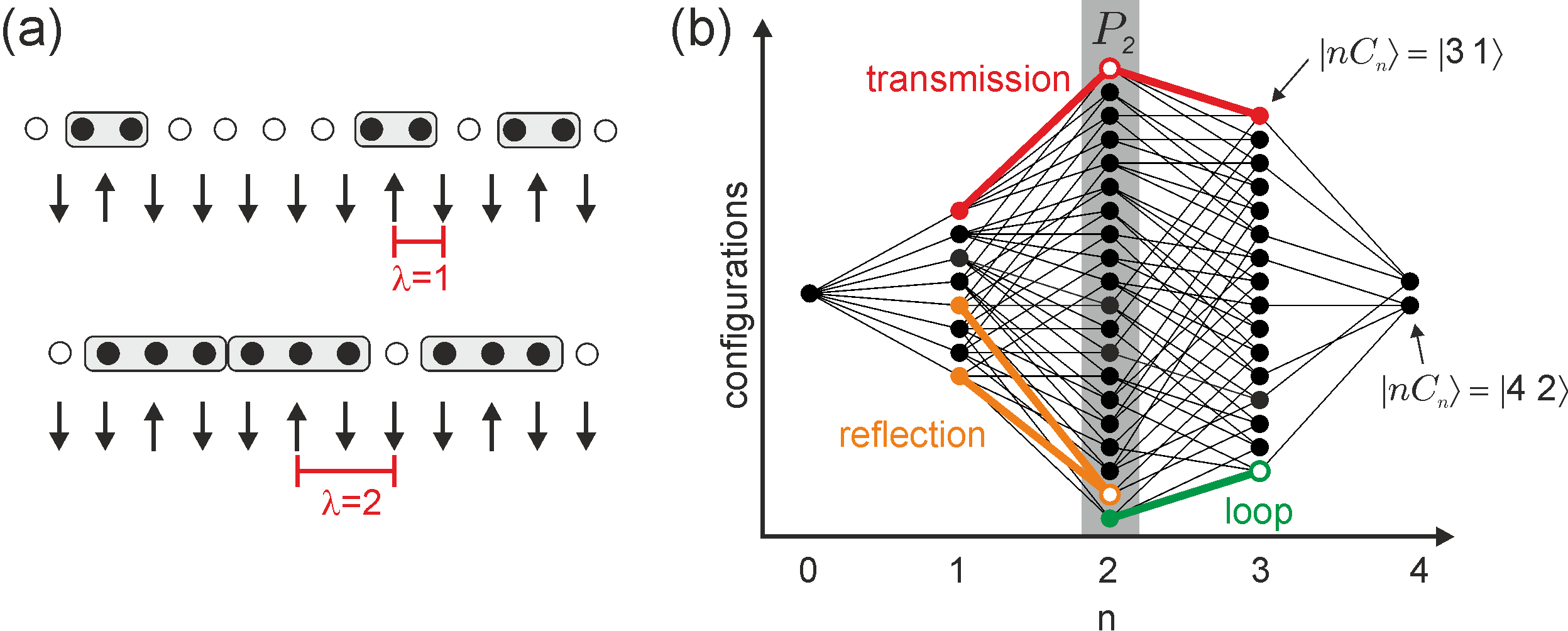}
\caption{(a) Representation of configurations in terms of spins (arrows) and the hard rods (grey rectangles). Each up spin inhibits further excitations on $\lambda$ neighboring sites on both its left and right. This blockade effect can equivalently be illustrated by hard rods on the lattice, each occupying $\lambda+1$ sites, and spin flips correspond to placing/removing hard rods on the lattice. (b) An example of the configuration network for $L=8$ and $\lambda =1$. Each node represents a classical arrangement of $n$ hard rods $| n \mathcal{C}_n \rangle$ (see text). The links between nodes connect configurations that can be converted into one another by the deposition/removal of a hard rod. Since we are interested in the dynamics in number space we do not resolve individual states but consider the time evolution of the operator $P_n$ that projects on all basis states contained in the $n$-th column. The equation of motion for $\langle P_n \rangle$ depends directly on the structure of the graph, specifically on the abundance of loops (green), reflections (orange) and transmissions (red).}
\label{fig:System}
\end{figure}

The Hilbert space of the system is spanned by all classical spin configurations compatible with the blockade condition. In \fref{fig:System}(a) we provide examples of permitted spin configurations for the cases $\lambda=1$ and $\lambda=2$. Instead of using spins we employ a description of the basis states in terms of hard rods. Each configuration of spins can be uniquely mapped into and arrangement of hard rods of length $\lambda +1$, as illustrated in \fref{fig:System}(a). A spin-flip from the down- to the up-state corresponds to the deposition a hard rod, while the opposite process removes a hard rod from the lattice. Our aim is to analyze the non-equilibrium dynamics of the hard rods effectuated by $H_{\mathrm{spin}}$ after a quench. In particular we seek an effective evolution equation for the probability density $p_n (t)$, which describes the probability of having the lattice occupied by $n$ hard rods at time $t$. The first step towards such equation is to gain an understanding of the Hilbert space structure. Here it is convenient to employ a graphical representation of the Hilbert space as a configuration network. We will see that the structure of the effective evolution equation for the $p_n$ will crucially depend on the statistical properties of the network. In the following, we will construct the network and analyze its properties in detail.

\section{Configuration network}\label{sec:config_network}
\subsection{Structure}
To construct the configuration network we introduce the basis states $| n \mathcal{C}_n \rangle$. Each of these states represents a specific classical arrangement $\mathcal{C}_n$ of $n$ hard rods. These states satisfy the completeness $\sum_{n\mathcal{C}_n}\left|n\mathcal{C}_n\right>\left<n\mathcal{C}_n\right|= \dblone$ and orthonormality relation $\left<n\mathcal{C}_n |m\mathcal{K}_m\right> = \delta_{nm}\delta_{\mathcal{C}_n\mathcal{K}_m}$. We interpret each of the microstates $| n \mathcal{C}_n \rangle$ as a node of a network. By grouping configurations containing the same number of hard rods into columns we obtain the network structure depicted in  \fref{fig:System}(b). The time-evolution of the system can then be imagined as a temporal change in the occupation of these nodes. Dynamics is introduced through the Hamiltonian $H_\Omega$ which leads to transitions between microstates that are represented as edges of the network. Since $H_\Omega$ causes only single spin flips, nodes in neighboring columns are only linked directly, if their corresponding microstates can be converted into one another by the removal/deposition of one hard rod. For example, setting $\lambda=1$ the state $\left|\downarrow\uparrow\downarrow\downarrow\uparrow\downarrow\downarrow\uparrow\downarrow\right>$ in the $n=3$ column is directly linked with $\left|\downarrow\uparrow\downarrow\downarrow\downarrow\downarrow\downarrow\uparrow\downarrow\right>$, but not with $\left|\downarrow\downarrow\uparrow\downarrow\downarrow\downarrow\uparrow\downarrow\right>$ in the $n=2$ column. In the following we will analyze in detail the properties of the configuration network.

\subsection{Properties of the configuration network}
The most basic properties that define the structure of our configuration network are the number of columns and the number of nodes within each column. Fixing the length of the system to $L$ sites, and applying periodic boundary condition, the maximum number of hard rods that each occupy $\lambda+1$ sites (i.e., the blockade radius is $\lambda$), which can be placed on the lattice is $\lfloor L/(\lambda+1) \rfloor$, where $\lfloor x \rfloor$ denotes the closest integer smaller or equal to $x$. The index counting the number of hard rods can thus take the values $n = 0,1,..., \lfloor L/(\lambda+1) \rfloor$. The number of microstates $\nu_{n}$ contained in the $n$-th column is given by the number of ways in which one can distribute $n$ indistinguishable hard rods of length $\lambda+1$ over $L$ lattice sites. This is a standard combinatorial problem with solution
\begin{equation}
\nu_{n} = \frac{ L (L-1-\lambda n)!}{n!(L-(\lambda+1)n)!} \, .
\end{equation}
Having determined the properties of the ``backbone'' of the network we now turn to assessing the linkage of the nodes. In particular, we calculate the mean number of different possibilities $T_{n\to n\pm 1}$ to go from a state with $n$ hard rods to one in the adjacent columns. This quantity can be expressed as $T_{n\to n\pm 1} = c_{n, n \pm 1}/\nu_n$, where $c_{n, n \pm 1}$ denotes the \emph{total} number of links between columns $n$ and $n \pm 1$ and hence  $c_{n \pm 1 , n} = c_{n , n \pm 1}$. (Note that this symmetry does not hold for the $T_{n \to m}$.) Moreover, we know that $T_{n \to n-1} = n$, as in a configuration of $n$ hard rods there are $n$ possibilities for removing one hard rod reaching a state with $n-1$ hard rods. Using these relations the total number of links between two columns evaluates to
\begin{eqnarray}
c_{n , n-1} &= &T_{n\to n-1}\nu_n =  \frac{ L (L-1-\lambda n)!}{(n-1)!(L-(\lambda+1)n)!}, \nonumber \\
c_{n+1, n } &= &T_{n+1\to n}\nu_{n+1} = \frac{ L (L-1-\lambda (n+1))!}{n!(L-(\lambda+1)(n+1))!},\nonumber \\
\end{eqnarray}
which can then be used to calculate $T_{n \to n+1}$.

\begin{figure}
\centering
\includegraphics[width=0.5\columnwidth]{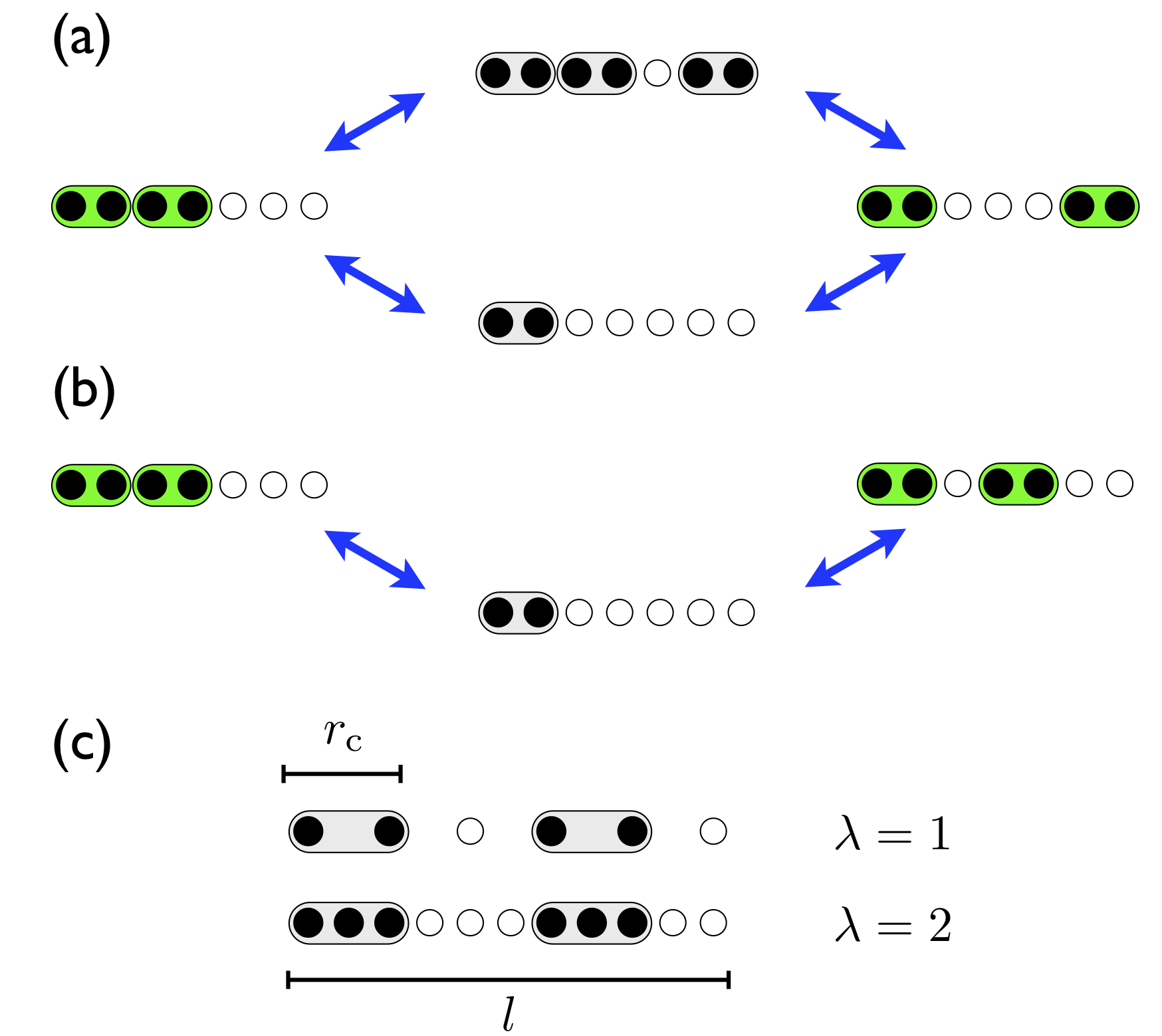}
\caption{(a,b) Graphical illustration of the reflection processes. Two configurations can at most be connected by two reflection pathways (a). However, since rods must not overlap, the path containing an additional hard rod during the intermediate step is often forbidden (b). (c) For the numerical study we fix the physical system length $l$ and the critical radius $r_\mathrm{c}$. The parameter $\lambda$ is varied by increasing the number of lattice sites.}\label{fig:reflections}
\end{figure}

Let us continue by analyzing the second order processes shown in \fref{fig:System}(b), i.e. loops, reflections and transmissions, as they will enter in the derivation of the Master equation. Selecting a node from the network, the number of loop transitions from that node equals the number of links that this node has with other nodes. Therefore, the mean number of loop transitions from a state with $n$ hard rods is given by
\begin{equation}
\overline{N_{\mathrm{loop}}^{(n)}} = T_{n\to n+1} + T_{n \to n-1}.
\label{eq:diagonal}
\end{equation}
Reflections connect two configurations that contain the same number of hard rods but differ in the position of exactly one hard rod [cf. \fref{fig:reflections} (a,b)]. Two configurations that are randomly selected from one column will typically differ by the positioning of more than one hard rod and are thus not connected by a reflection. If two microstates happen to be connected there are at most two paths as illustrated in \fref{fig:reflections}: a deexcitation followed by an excitation or vice versa. For $n \gg 1$ (at high density) there is often even only one path available [\fref{fig:reflections}(b)]. Similar considerations can also be made for transmission diagrams, i.e., the average number of paths connecting two microstates containing $n$ and $n\pm2$ hard rods is $\sim 1$. In fact the mean number of  reflections (transmissions) $\overline{N_{\mathrm{refl}}^{(n)}}$ ($\overline{N_{\mathrm{trans}}^{(n)}}$) between two \emph{randomly} selected states can be calculated analytically:
\begin{eqnarray}
\overline{N_{\mathrm{refl}}^{(n)}} &=& \frac{T_{n+1 \to n}(T_{n+1\to n}-1)\nu_{n+1}}{\nu_n(\nu_n-1)} +\frac{T_{n-1 \to n}(T_{n-1\to n}-1)\nu_{n-1}}{\nu_n(\nu_n-1)}, \nonumber\\
\overline{N_{\mathrm{trans}}^{(n)}} &= & \frac{T_{n+2 \to n+1}T_{n+1\to n}+T_{n-2\to n-1}T_{n-1\to n}}{\nu_n}.\nonumber \\
\label{eqn:average_transitions}
\end{eqnarray}
To illustrate that loop transitions are far more abundant than reflections and transmissions we present some numerical examples in \tref{tab:loop}. Here we compare $\overline{N_{\mathrm{loop}}^{(n)}}$, $\overline{N_{\mathrm{refl}}^{(n)}}$ and $\overline{N_{\mathrm{trans}}^{(n)}}$ for a number of lattice and hard rod sizes. This leads to two observations. First, the relative weight of loop transitions largely increases with increasing systems size $L$, as due to the larger dimension of the Hilbert space each state can have more connections to other configurations. Second, the probability that two states within a column of the network are connected by a reflection is vanishingly small in the ''bulk'' of the network ($1 \ll n \ll \lfloor L/(\lambda+1) \rfloor$). The same is also true for transmissions between columns $n$ and $n \pm 2$. Note, that near the boundaries of the network, i.e. columns close to the maximum/minimum $n$-value, the condition $\overline{N_{\mathrm{loop}}^{(n)}}\gg \overline{N_{\mathrm{trans}}^{(n)}},\overline{N_{\mathrm{refl}}^{(n)}}$ are less well satisfied. However, this concerns only an exponentially small subset of states forming the configuration network. These two observations on the statistics of the configuration network are of central importance in the derivation of the effective Master equation for $p_n(t)$ that we are going to present in the following section.

\begin{table}
\centering
  \begin{tabular}{|c |c|c|| c| c|c|}
  \hline
     L& $\lambda$ & $n$ & $ \overline{N_{\mathrm{loop}}^{(n)}}$ & $\overline{N_{\mathrm{refl}}^{(n)}}$ & $\overline{N_{\mathrm{trans}}^{(n)}}$  \\ \hline
     & & 2 & 8.22 & 0.54 & 2.22\\
     \cline{3-6}
     \raisebox{1ex}[0pt]{12} &  \raisebox{1ex}[0pt]{1} & 4 & 5.71 & 0.17& 0.51 \\
     \hline
      &   & 2 & 296 & 0.026& $2$ \\ \cline{3-6}
      &  \raisebox{1ex}[0pt]{1}  & 75 & 174.78 & $1.27\times10^{-57}$& $1.39\times10^{-57}$ \\ \cline{2-6}
      &   & 2 & 280.1 & 0.026 & $2$\\ \cline{3-6}
      &  \raisebox{1ex}[0pt]{5}  & 25 & 93.39 & $1.64\times10^{-27}$& $3.18\times10^{-27}$ \\ \cline{2-6}
       &   & 2 & 244.78 & 0.025&$2$ \\ \cline{3-6}
      \raisebox{7ex}[0pt]{300}  &  \raisebox{1ex}[0pt]
      {14}  & 10 & 68& $3.11\times10^{-13}$&$ 1.73\times10^{-12}$ \\      \hline
  \end{tabular}
  \caption{Average number of connection between nodes established by the three types of second order processes depicted in \fref{fig:System}(b). We have chosen a relatively small lattice size ($L=12$) and a large one ($L=300$) for illustration.}   \label{tab:loop}
\end{table}

\section{Master Equation}\label{sec:master_equation}
In this section we derive an equation of motion which describes the evolution of number of hard rods on the lattice as function of time. This means that we are not interested in the actual population of individual nodes of the graph shown in \fref{fig:System}(b) but rather in the probability $p_n(t)$ of the system to reside in a specific column $n$ at time $t$. We will see that this eventually leads to a Master equation that has a steady state $p_n(t \to \infty)$ which is proportional to the number of classical configurations contained in a specific column. This strongly suggests that this steady state corresponds to a microcanonical equilibrium state in which all arrangements of hard rods occur with equal probability.

\subsection{From the von-Neumann equation to the Master equation}\label{sec:master_equation_deriv}
The probability $p_n$ is defined as $p_n=\Tr \rho(t) P_n$, where $\rho(t)$ is the density matrix of the system and $P_n = \sum_{\mathcal{C}_n} \left|n\mathcal{C}_n\right>\left<n\mathcal{C}_n\right|$ is a projector which projects onto the subspace spanned by all microstates contained in the $n$-th column of the configuration network [see \fref{fig:System}(b)]. Throughout this work we are interested in a situation where the initial state of the system $\rho(0)$ contains a fixed number of hard rods, i.e. $\left[P_n,\rho(0)\right]=0$.

To begin with the formal derivation of the Master equation, let us momentarily return to the case of finite $V$ and transform the Hamiltonian (\ref{eqn:spin_H}) into the interaction picture with respect to $H_V$. As shown in \ref{sec:integrated_v_N_equation} one can then write the evolution equation for $p_n = \left<P_n\right> = \Tr \rho(t)P_n$ as
\begin{eqnarray}
\partial_t {{\left<P_n\right>}}_t &=&   -\int_0^t ds\, \Tr \{ P_n H_I(t) H_I(s) \rho(s)    +H_I(s)H_I(t)P_n\rho(s)
\nonumber \\ &&- H_I(s)P_n H_I(t)  \rho(s)  -H_I(t)P_n H_I(s) \rho(s)\},
\label{eqn:integral_von-Neumann_expand}
\end{eqnarray}
with
\begin{eqnarray}
H_I(t) &=& \sum_{n\mathcal{C}_n, m\mathcal{K}_m} \left<n\mathcal{C}_n|H_\Omega|m\mathcal{K}_m\right> e^{-i(\omega_{m\mathcal{K}_m}-\omega_{n\mathcal{C}_n})t}\left|n\mathcal{C}_n \right> \left<m\mathcal{K}_m\right|,
\label{eqn:inter_pic}
\end{eqnarray}
where $\omega_{n\mathcal{C}_n}=\left<n\mathcal{C}_n|H_V|n\mathcal{C}_n\right>$ denotes the configuration energy of the spin configuration $|n\mathcal{C}_n \rangle$. In order to evaluate the integral in the still exact evolution \eref{eqn:integral_von-Neumann_expand} we make the replacement  $\rho(s)\to\rho(t)$, which effectively amounts to a second order approximation. At this point the quality of this approximation is not clear but we will provide a numerical justification of this step \emph{a posteriori}.
Returning to the ideal blockade regime ($V/\Omega \to \infty$) the fact that each allowed spin configuration has the same configuration energy, $\omega_{n\mathcal{C}_n} = 0$, removes the exponential factor of the interaction picture Hamiltonian making it time-independent ($H_I(t) \to H_{\Omega}$). Thus, the time-integration in \eref{eqn:integral_von-Neumann_expand} can be carried out and simply amounts to a multiplication by $t$. The steps result in the second order equation
\begin{equation}
\partial_t {{\left<P_n\right>}}_t = -t \mathrm{Tr} \left[
P_n H^2_{\Omega} + H^2_{\Omega} P_n - 2 H_{\Omega} P_n H_{\Omega}
\right] \rho(t),
\label{eq:after_integration}
\end{equation}
that depends on products of the form
\begin{equation}
\sum_ {m\mathcal{K}_m } \left< n\mathcal{C}_n|H_\Omega |m\mathcal{K}_m\right>\left< m\mathcal{K}_m|H_\Omega |p\mathcal{L}_p\right> = \left<n\mathcal{C}_n\right| H_\Omega^2 \left|p\mathcal{L}_p\right>  \nonumber
\label{eqn:matrix_elements}.
\end{equation}
These matrix elements are the second order processes discussed in the previous section. Diagonal elements correspond to loops and off-diagonal ones to either reflections or transmission. From the analysis of the graph we know that loop transitions are far dominant and that $H_\Omega^2$ is thus approximately diagonal, hence
\[
\left<n\mathcal{C}_n\right| H_\Omega^2 \left|p\mathcal{L}_p\right> \approx
\left<n\mathcal{C}_n\right| H_\Omega^2 \left|n\mathcal{C}_n\right> \delta_{np} \delta_{\mathcal{C}_n \mathcal{L}_p}.
\]
The final approximation consists of neglecting the variation of the matrix elements of $H^2_{\Omega}$ with $\mathcal{C}_n$ within a given column of the configuration network, by replacing the matrix element with its average taken over all microstates within a column
\begin{equation}
\left<n\mathcal{C}_n\right| H_\Omega^2 \left|n\mathcal{C}_n\right> \to
\Omega^2 \overline{N_{\mathrm{loop}}^{(n)}} =
\Omega^2 \left( T_{n \to n + 1} + T_{n  \to n-1 } \right).
\end{equation}
Inserting these approximations into \eref{eq:after_integration} and writing $p_n (t) = \mathrm{Tr} P_n \rho(t) = \sum_{\mathcal{C}_n} \mathrm{Tr} \left|n\mathcal{C}_n\right> \left<n\mathcal{C}_n\right| \rho(t)$ we arrive at the desired Master equation for the dynamics of the probability distribution $p_n$ of finding $n$ hard rods in the system at time $t$
\begin{eqnarray}
\partial_t {p}_{n} \left(t\right) &=& 2{\Omega}^2 t \left[ T_{n+1\rightarrow n} p_{n+1}(t) + T_{n-1\rightarrow n} p_{n-1}(t)\right] \nonumber \\
 &&- 2{\Omega}^2 t \left[ T_{n\rightarrow n-1}  + T_{n\rightarrow n+1} \right] p_{n}(t) .
\label{eqn:master_eqn}
\end{eqnarray}
This Master equation has a stationary solution despite the explicit appearance of the time variable on the right hand side. This can be seen by transforming to a new time variable $\tau = \Omega t^2$ which removes the explicit time-dependence. The rate coefficients of the Master equation (\ref{eqn:master_eqn}) obey detailed balance, i.e., $p_n^{\mathrm{eq}}T_{n \to n+1} = p_{n+1}^{\mathrm{eq}}T_{n+1 \to n}$, where  $p_n^{\mathrm{eq}}$ denotes the steady state distribution
\begin{equation}
p_n^{\mathrm{eq}}= \frac{\nu_n}{\sum_{n=0}^{L/(\lambda+1)}\nu_n}.
\label{eqn:counting_dist}
\end{equation}
This distribution is proportional to the number $\nu_n$ of arrangements of $n$ hard rods, suggesting that each node of the network is populated with equal probability, as one would expect from a "maximum entropy state".

As shown in reference \cite{atga+:12}, \eref{eqn:master_eqn} is a discretized Fokker-Planck equation in excitation number space. Its drift coefficient is proportional to $T_{n\rightarrow n-1}  - T_{n\rightarrow n+1}$ and its diffusion coefficient proportional to $T_{n\rightarrow n-1}  + T_{n\rightarrow n+1}$. The Master equation thus describes the excitation dynamics of the system as diffusion between the columns of the configuration network shown in \fref{fig:System}(b).

\subsection{Time Evolution and Steady State of the Master Equation} \label{sec:time_ev_and_steady_state}
\begin{figure}
\centering
\includegraphics[width=0.5\columnwidth]{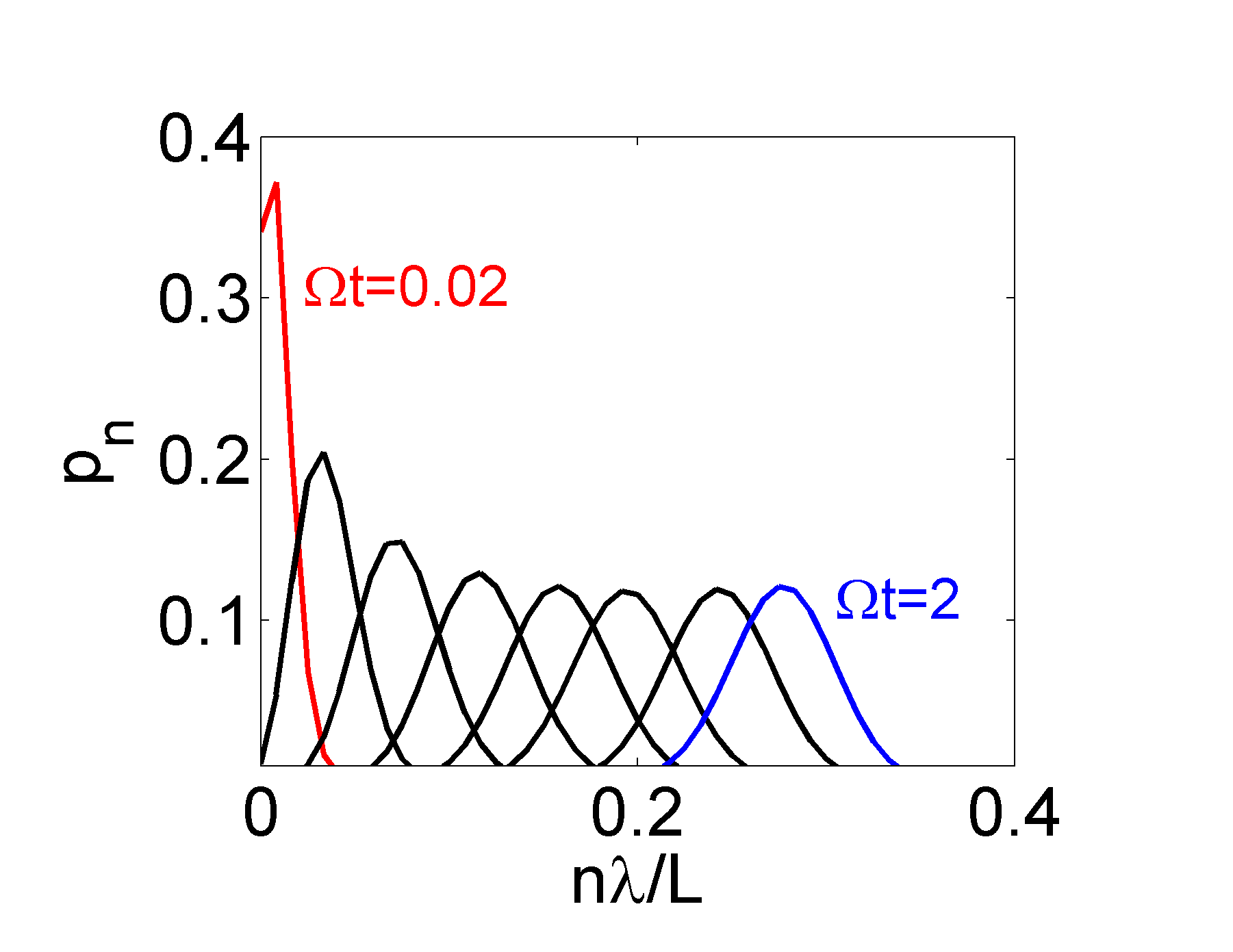}
\caption{Time evolution of the distribution function $p_n(t)$ as a function of the hard rod density $ n\lambda /L$ for $\lambda=1$ (nearest neighbor blockade). The system considered has a lattice length to blockade radius ratio $l/r_{\mathrm{c}}=120$. From left to right $p_n$ is shown at: $\Omega t=0.02 \mathrm{ (red)},0.1,0.2,0.3,0.4,0.5,0.6,0.8$ and $\Omega t=2$ (blue, corresponding to the steady state distribution). The initial distribution function used is $p_0 = \delta_{n,0}$.
}
\label{fig:Master_n.n.}
\end{figure}

\begin{figure}
\includegraphics[width=1\columnwidth]{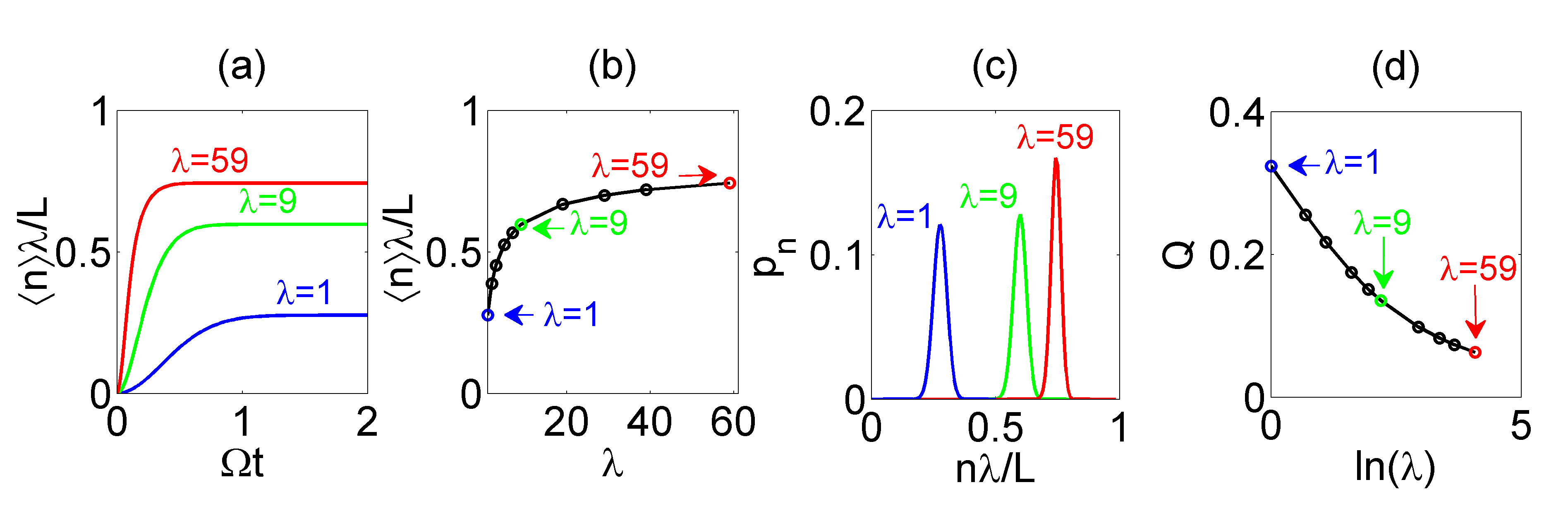}
\caption{
Dependence of the time evolution and steady state properties on $\lambda$.
(a) Time-evolution of the average hard rod density $ \langle n \rangle \lambda /L$ starting from the configuration without hard rods for different  numbers of blockaded sites: $\lambda =1$ (blue), $\lambda =9$ (green) and $\lambda =59$ (red). (b) Steady state values of $\langle n \rangle \lambda /L$ as function of $\lambda$. (c) Steady state distribution $p_n$ as function of the hard rod density $ n \lambda /L$ for the three values of the number of blockaded sites used in (a). (d) Steady state value of the parameter Q as function of the number of blockaded sites. For more details, see text.
}
\label{fig:Master_diff_lambda}
\end{figure}

We will now study the time evolution of the distribution function $p_n$ by numerically solving \eref{eqn:master_eqn}. We perform the following analysis by fixing both the physical lattice length $l$ and the critical radius $r_{\mathrm{c}}$. The ratio $l/r_{\mathrm{c}}$ is then also fixed. We choose a value of $l/r_{\mathrm{c}}=120$ in the numerical examples of this subsection, which means that at most $120$ hard rods can be placed on the lattice. What we change is the number of lattice sites occupied by a hard rod, i.e. $\lambda+1$, which is done by varying the total number $L$ of lattice sites. This is illustrated in \fref{fig:reflections}(c). We note that for $\lambda \to \infty$ this procedure results in taking the continuum limit $a/r_{\mathrm{c}} \to 0$.

\Fref{fig:Master_n.n.} shows the time evolution of $p_n$ as a function of the hard rod density $n\lambda /L$ for $\lambda=1$ starting from the initial state with no hard rods, i.e., $p_n (0) = \delta_{n,0}$. The time evolution of the distribution function towards its steady state (indicated in blue) shows the characteristic features expected from a system following a Fokker-Planck evolution: $p_n$ gets broader due to diffusion between the columns of the configuration network and drifts towards its equilibrium position, where the rate coefficients for further excitation ($T_{n\to n+1}$) and de-excitation ($T_{n\to n-1}$) become equal.

\Fref{fig:Master_diff_lambda} summarizes the behavior of the solution of the Master equation for varying $\lambda$. Panel (a) shows the time evolution of the mean hard rod density $ \langle n \rangle \lambda /L$ starting from an empty lattice. For short times the hard rod density exhibits a quadratic time dependence, $ \langle n \rangle \lambda /L \propto t^2$. This behavior is induced by the explicit linear time dependence of the right hand side of the Master equation. For long times $ \langle n \rangle \lambda /L$ saturates to its steady state value, which is reached faster the larger $\lambda$. Furthermore, the hard rod density in the steady state gets larger as $\lambda$ increases [\fref{fig:Master_diff_lambda}(b)]. Interestingly, however, the fluctuations around the mean density $ \langle n \rangle \lambda /L$ decrease with increasing $\lambda$, which can be seen in \fref{fig:Master_diff_lambda}(c), where the full steady state distribution function $p_n$ is depicted for three different values of $\lambda$. This behavior is further quantified in \fref{fig:Master_diff_lambda}(d), where we show the steady state value of the ratio $Q \equiv (\langle n^2 \rangle - \langle n \rangle^2)/\langle n \rangle$, which relates the width of the distribution function to its mean, as a function of $\lambda$. The ratio asymptotically approaches zero with increasing $\lambda$, i.e. increasing number of lattice sites occupied by a hard rod [cf.\ \fref{fig:Master_diff_lambda}(d)]. This surprising feature is the result of an entropic effect which has recently been discussed in reference \cite{atle:12}.

\section{Comparison with the Exact Quantum Evolution}\label{sec:numerics}

\begin{figure}
\centering
\includegraphics[width=1\columnwidth]{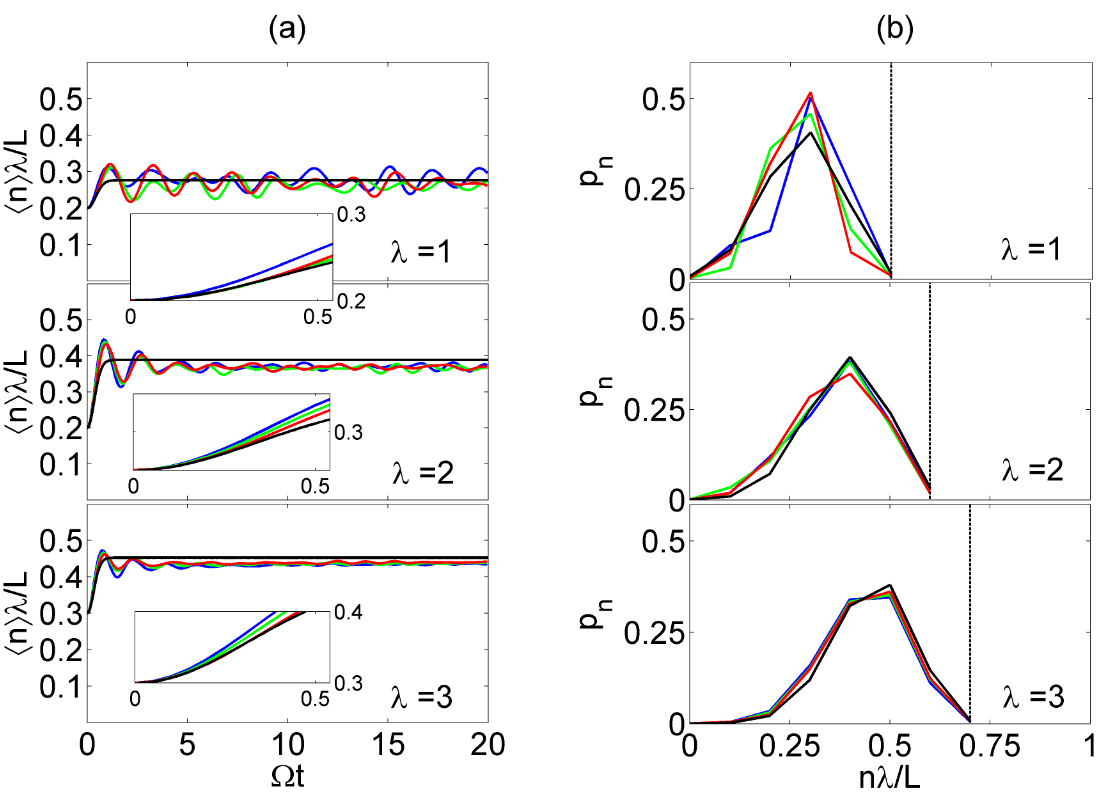}
\caption{
Comparison of the exact quantum dynamics with the predictions of the Master equation (\ref{eqn:master_eqn}) for different $\lambda$. (a) Time evolution of the hard rod density and (b) the distribution functions at $\Omega t =20$.  The vertical dashed lines indicate the maximum possible hard rod density for each $\lambda$, which is given by $\lambda/(\lambda+1)$. From top to bottom, the blockade radius is increased as $\lambda = 1,2,3$ while the ratio $l/r_{\mathrm{c}}=10$ remains fixed.  In all plots, the black curves are the solutions of the Master equation. The colored curves show the solution of the Schr\"odinger equation using three randomly chosen initial states $| n_0 \mathcal{C}_{n_0} \rangle$ with the same number $n_0$ of initial hard rods (from top to bottom $n_0 = 2,2,3$).
}
\label{fig:blockade_dependence}
\end{figure}

After having discussed the main features of the time evolution and the steady state of the Master equation we will now compare its predictions to the exact quantum dynamics of the system. To this end we have numerically solved  Sch\"odinger's equation with the Hamiltonian (\ref{eqn:spin_H}) in the limit of $V/\Omega \to \infty$ for up to $L=30$ sites. The left column of \fref{fig:blockade_dependence} shows the numerically exact quantum evolution of the hard rod density together with the prediction of the Master equation (in black). The ratio of system length to critical radius is fixed to  $l/r_{\mathrm{c}}=10$ and we choose $\lambda = 1, 2, 3$ from the top to the bottom panel. Note that the dimension of the Hilbert space increases from top to bottom. The differently colored curves in each panel show solutions to the Schr\"odinger equation starting from randomly chosen initial states $| n_0 \mathcal{C}_{n_0} \rangle$ with the same number of hard rods $n_0$ but different spin configurations $\mathcal{C}_{n_0}$. The initial number of hard rods is chosen such that the initial state lies in a region of the configuration network with large connectivity, i.e., where the statistical assumptions underlying the derivation of the Master equation are well met. The right column shows the corresponding probability distributions $p_n$ at $\Omega t =20$. For both $ \langle n \rangle \lambda /L$ and $p_n$ the agreement between the results of the exact quantum calculation and the prediction of the Master equation is remarkably good. In particular, for long times the results of the full quantum calculation and solution of the Master equation only differ by roughly one per cent. For short times the quadratic time dependence of $ \langle n \rangle \lambda /L$ as well as its dependence on $\lambda$ are well reproduced (see insets). For longer times the full quantum solutions exhibit oscillations around the equilibrium value of $ \langle n \rangle \lambda /L$. Being a simple rate equation our Master equation does not reproduce these. However, the quantum oscillations decrease with increasing dimension of the Hilbert space. For $\lambda =1$  this behavior was also reported in reference \cite{atga+:12}.

\begin{figure}
\centering
\includegraphics[width=1\columnwidth]{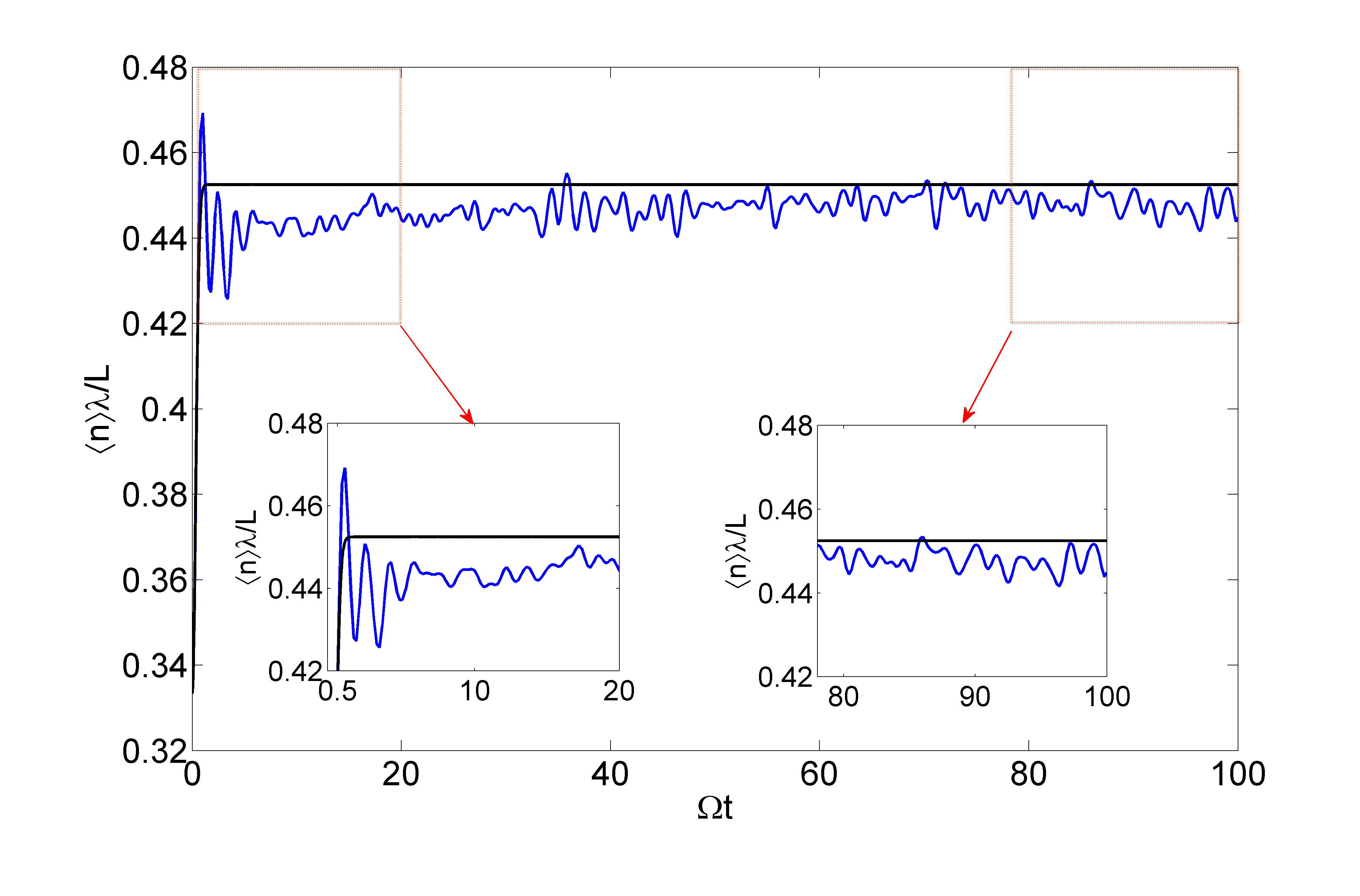}
\caption{Long time quantum evolution of the hard rod density starting from a randomly chosen initial state with $n_0 =3$ hard rods for $\lambda = 4$ and $l/r_{\mathrm{c}} = 9$. The evolution predicted by the Master equation is plotted in black. To highlight the shift, the time evolution from $\Omega t=0.5$ to $\Omega t=20$ is enlarged and shown in the left inset, and the evolution from $\Omega t=80$ to $\Omega t=100$ is enlarged and shown in the right inset.}
\label{fig:long_time_evolve}
\end{figure}

At long times we observe a small systematic offset of $ \langle n \rangle \lambda /L$ obtained from the exact numerics from the steady state values predicted by the Master equation. In the cases shown in \fref{fig:blockade_dependence}(a), where the initial state contains fewer hard rods than the equilibrium state, the quantum calculations suggest a sightly lower value of the average hard rod density at long times. In contrast, for initial states with a higher hard rod density than the equilibrium value the quantum results lie sightly above the prediction of the Master equation. Due to this systematic dependence on the initial state, we attribute this small offset to the presence of memory effects in the quantum dynamics that were completely neglected in the derivation of the Master equation. Furthermore, the data of the full quantum calculation for $\lambda=3$  exhibit a very slow drift of $ \langle n \rangle \lambda /L$ towards the steady state of our rate equation. This effect is seen more clearly in \fref{fig:long_time_evolve}, where we follow the time evolution of the hard rod density for $\lambda =4$ to much longer times. The observed shift might be indicative of a pre-equilibration process in the quantum system, in which $\langle n \rangle \lambda /L$ quickly reaches a quasi-equilibrium state, which then very slowly equilibrates to the ``true'' steady state. However, the drift might also stem from a long wavelength oscillation present in the full quantum calculation due to the finite size of the system. Since the full quantum calculations are limited to small system sizes it is difficult to further explore this effect, which seems to be more pronounced with increasing $\lambda$.

Let us finally return to the observation made in \fref{fig:blockade_dependence}(b) that distribution functions $p_n$ calculated fully quantum mechanically and using the Master equation agree very well at long times. In order to quantify the degree of agreement  we use the following overlap measure \cite{bezy:06},
\begin{equation}
\mathcal{D}=1-\frac{1}{2}\sum_{n=0}^{L/(\lambda+1)} |\bar{p}_n-p_n^{\mathrm{eq}}| \, .\label{eq:overlap}
\end{equation}
Here $p_n^{\mathrm{eq}}$  denotes the steady state solution of the Master equation and $\bar{p}_n$ is the equilibrium distribution obtained from the solution of the Schr\"odinger equation, time averaged over a time interval $\Delta t$, i.e.,  $\bar{p}_n = \int_{t}^{t+\Delta t} \mathrm{d} \tau \, p_n (\tau) / \Delta t$. The distribution functions are identical when $\mathcal{D}=1$, and  $\mathcal{D}=0$ for distribution functions that are completely non-overlapping. For the three situations discussed in \fref{fig:blockade_dependence} we have selected $100$ randomly chosen initial spin configurations for $L=10, \lambda=1$ and $500$ randomly chosen initial spin configurations for $L=20, \lambda=2$ and $L=30, \lambda=3$, and have used them as initial states of the quantum evolution. The time average in order to compute $\bar{p}_n$ was taken over the interval $\Delta t = [20/\Omega, 40/\Omega]$.
\begin{figure}
\centering
\includegraphics[width=0.9\columnwidth]{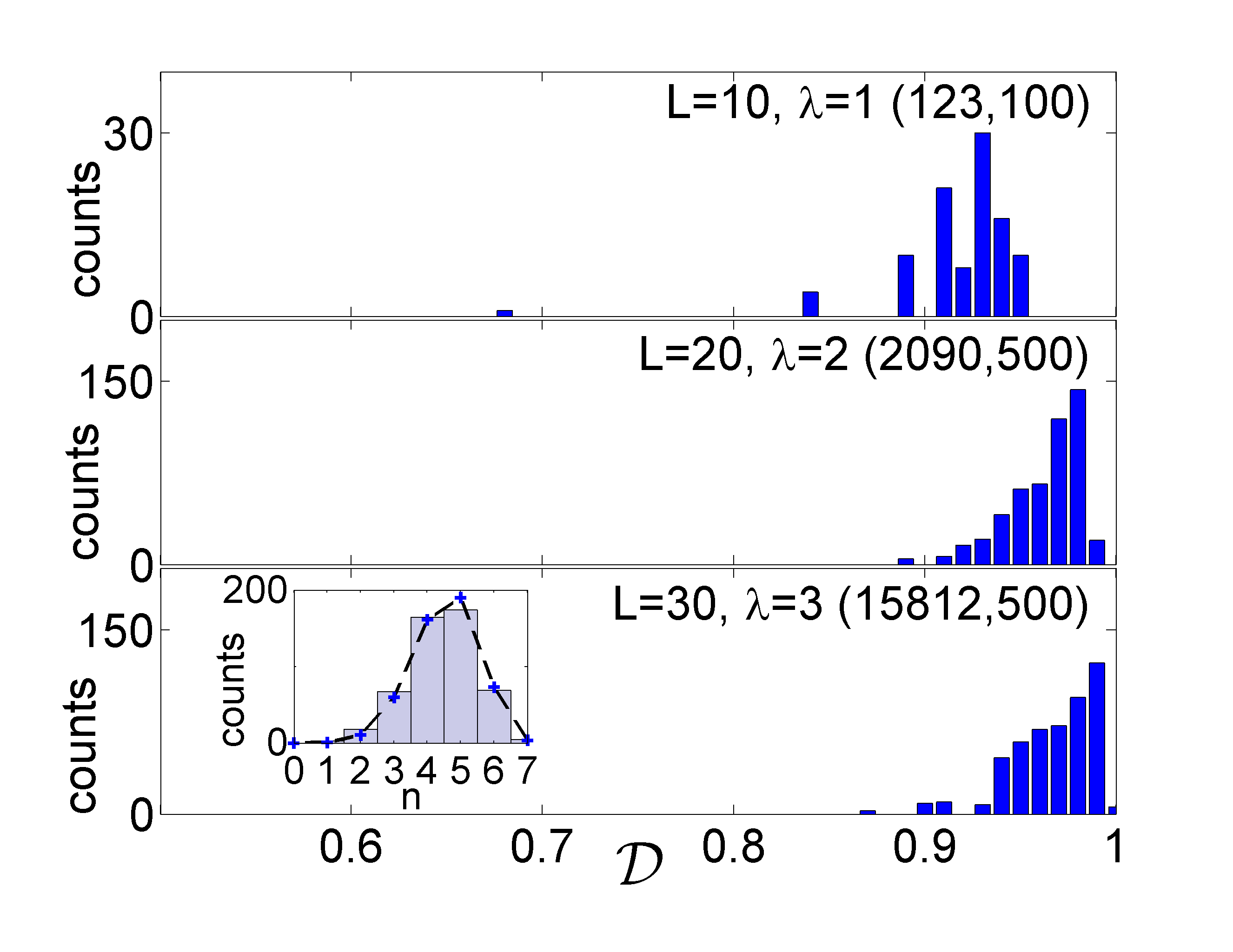}
\caption{
Histogram of the parameter $\mathcal{D}$, defined in \eref{eq:overlap}, for the three parameter sets used in \fref{fig:blockade_dependence}. The numbers in the parentheses show the dimension of the Hilbert space and the number of initial states used to create the histogram, respectively. For the bottom panel we also show the distribution of the number of hard rods contained in the initial states as an inset. The black dashed line with blue cross marks in the inset shows the microcanonical steady state distribution of the number of hard rods for comparison.
}
\label{fig:histogram}
\end{figure}
We have collected the results of these simulations in the histograms shown in \fref{fig:histogram}. Here we see that for the vast majority of initial conditions $\mathcal{D}$ is close to one with only a few outliers. This indicates that indeed equilibration is largely independent of the initial state.
In order to demonstrate that the used set of initial conditions is representative we have looked at the distribution of the number of hard rods contained in the initial states. As an example the inset in \fref{fig:histogram} shows this distribution for the parameters of the bottom panel. Comparing this to the microcanonical steady state distribution we see that indeed each number is represented with the correct weight. Finally, looking at the histograms shown in \fref{fig:histogram} as a function of the dimension of the Hilbert space we see that the system equilibrates better with increasing number of available microstates, confirming similar observations made in \cite{atga+:12} for the case $\lambda=1$. In the thermodynamic limit  the Master equation (\ref{eqn:master_eqn}) therefore indeed provides an excellent description of the non-equilibrium dynamics in particle number space.

\section{Conclusions and Outlook}\label{sec:outlook}
\begin{figure}
\centering
\includegraphics[scale = 0.7]{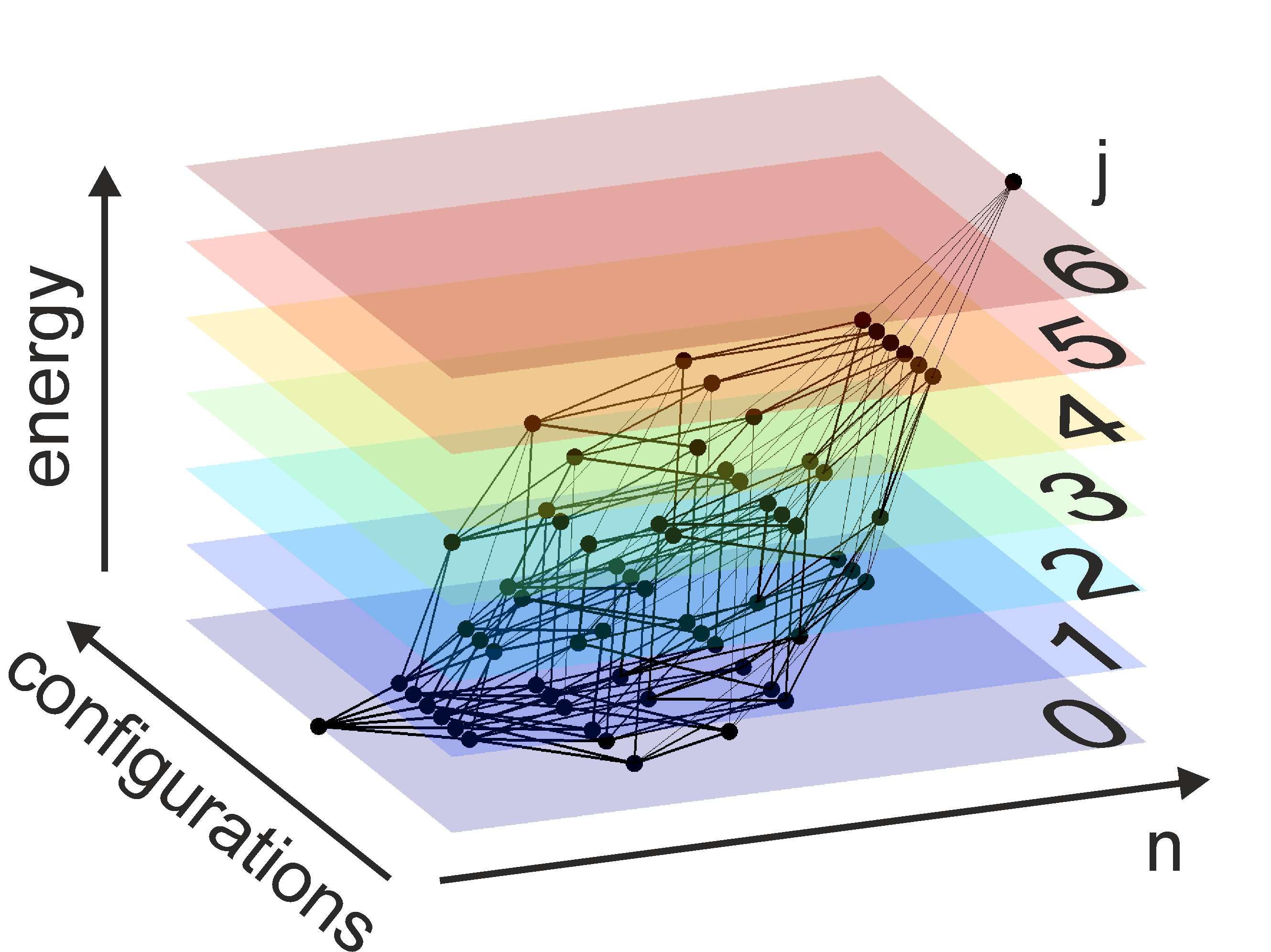}
\caption{Configuration network of a system with
$L=6$ sites, $\lambda=1$ and finite ratio $V/\Omega$. Different layers
represent different interaction energies and adjacent layers
are separated by $V$. The lowest layer corresponds to a network
similar to the one shown in \fref{fig:System}(b). Due to the structure of
the Hamiltonian transitions take place only between adjacent layers and
layers that are separated by $2V$. The timescale for
inter-layer-transitions is $\sim \Omega^2/V$ while
intra-layer-relaxation takes place on a time $\sim
\Omega^{-1}$. The index $j$ labels the number of adjacent nearest neighbor excitation pairs.} \label{figure3}
\end{figure}

In this work we have investigated the non-equilibrium dynamics and equilibration of a strongly interacting closed quantum system of hard rods of varying size. We showed that the steady state and the non-equilibrium relaxation towards it are well captured by a Master equation whose numerical coefficients can be analytically calculated using the combinatorics of hard rods placed on a lattice. The dependence of the relaxation dynamics on the initial state has been systematically investigated through numerical simulations of the full quantum dynamics. It has been found that relaxation into the steady state is achieved for the majority of initial conditions. Expectation values taken in the steady state are compatible with the assumption of an equal population of each microstate, suggesting that the system here is indeed well described by a microcanonical ensemble.

To conclude, let us finally comment qualitatively on large but finite interaction $V$, i.e. where the system no longer can be described in terms of hard rods. In this more general case we can still employ a representation of the state space in terms of a graph, but as shown in \fref{figure3} (for $\lambda=1$) a third dimension has to be added which accounts for the interaction energy $V$ of different combinations. The lowest layer corresponds to the set of nodes in the $V/\Omega\rightarrow\infty$ limit centered around zero energy with no contiguous excitations, cf. \fref{fig:System}(b). All higher layers contain $j$ pairs of adjacent excitations and hence are centered around the energy $\epsilon_j=j\times V$. Each of these layers can be thought of as an energy shell for which an intra-layer diffusive Fokker-Planck dynamics can be derived.  In reference \cite{leol+:10} it was numerically shown that the steady state population of these energy shells follows a Boltzmann law $p_j=p_{\epsilon_j} \propto e^{-\beta j V}$, with an inverse temperature $\beta$, provided that $V\gg\Omega$. This can be understood as follows: Choosing an initial state in the lowest energy shell the diffusive dynamics establishes a `maximum entropy state' within this shell on a timescale $\Omega^{-1}$. Transitions to higher energy shells take place on a slow timescale $\sim\Omega^2/V$. Single spin flips as effectuated by the Hamiltonian can, however, only couple energy shells with a distance of at most $2V$, i.e. a single spin flip can at most create two pairs of adjacent excitations. That means that the population of $j$-th energy shell can be estimated by $(\Omega^2/V^2)^j$ which approximates a Boltzmann distribution with temperature such that $\beta\Omega\approx -2(\Omega/V)\log (\Omega/V)$. It remains an intriguing question whether also in this more complex system the derivation of a Master equation is possible from first principles.

\ack
We wish to acknowledge discussions with J. Gaskell. C.A. acknowledges support through a Feodor-Lynen Fellowship of the Alexander von Humboldt Foundation. This work was funded in part by EPSRC Grant no. EP/I017828/1 and Leverhulme Trust grant no. F/00114/BG.

\appendix
\section{Integrated von-Neumann equation}\label{sec:integrated_v_N_equation}
In order to keep the notation compact we abbreviate our basis states as $\left|\alpha \right> = \left| n \mathcal{C}_n \right>$. The Hamiltonian given in \eref{eqn:spin_H} can then be expressed in this basis as
\begin{eqnarray}
  H = \sum_{\alpha}\left<\alpha|H|\alpha\right>\left|\alpha \right>\!\left<\alpha\right|+\sum_{\alpha, \beta \atop \alpha \ne \beta}\left<\alpha|H|\beta\right>\left|\alpha \right>\!\left<\beta\right| = H_V + H_{\Omega}.
\end{eqnarray}
We proceed by transforming the Hamiltonian into an interaction picture with respect to $H_V$ by applying a unitary transformation $\hat{U}= \exp[{-itH_V}]$ where we have set $\hbar =1$. The interaction picture Hamiltonian which is given by $H_I(t)=\hat{U}^\dagger H \hat{U}$, takes the following form:
\begin{equation}
H_I(t) = \sum_{\alpha, \beta} h_{\alpha,\beta}e^{-i(\omega_\beta-\omega_\alpha)t}\left|\alpha \right> \left<\beta\right|,
\end{equation}
where $\omega_\alpha=\left<\alpha|H_V|\alpha\right>$ and $h_{\alpha,\beta}=\left<\alpha|H_{\Omega}|\beta\right>$ are the diagonal and off-diagonal entries of the Hamiltonian, respectively. The density matrix $\rho(t)$ of the system evolves (in the interaction picture) according to the von-Neumann equation
\begin{equation}
\partial_t\rho(t) =  -i[H_I(t),\rho(t)].
\label{eqn:von-Neumann}
\end{equation}
We proceed by formally integrating the von-Neumann equation, finding that,
\begin{equation}
\rho(t) =  \rho(0)  - i\int_0^t ds [H_I(s),\rho(s)].
\label{eqn:von-Neumann-itegral}
\end{equation}
Substituting this integrated Eqn. (\ref{eqn:von-Neumann-itegral}) back into the von-Neumann equation (\ref{eqn:von-Neumann}), one obtains an integral form of the von-Neumann equation which reads
\begin{equation}
\dot{\rho}(t) =  -i[H_I(t), \rho(0)]- \int_0^t ds [H_I(t),[H_I(s),\rho(s)]].
\label{eqn:integral_von-Neumann}
\end{equation}
For an observable $\hat{O}$, the expectation value at a given time can be expressed as, $\hat{\left< O \right>}_t = \Tr \rho(t) \hat{O}$. Therefore, using the integral form of the von-Neumann equation, the rate of change of  $\hat{O}$ can be written as,
\begin{equation}
\partial_t {\hat{\left<O\right>}}_t = -\int_0^t ds\, \Tr \{ \hat{O}[H_I(t),[H_I(s), \rho(s)]]\}.
\end{equation}
Here we have assumed that the initial density matrix $\rho(0)$ commutes with the observable $\hat{O}$. Then, the first term in Eqn. (\ref{eqn:integral_von-Neumann}) yields a zero value when performing the trace. Expanding the double commutators and using the cyclic property of the trace to interchange density matrix to the rightmost side of each term, we find
\begin{eqnarray}
\partial_t {\hat{\left<O\right>}}_t &=&   -\int_0^t ds\, \Tr \{ \hat{O} H_I(t) H_I(s) \rho(s)   +H_I(s)H_I(t)\hat{O}\rho(s) \nonumber \\
&&
- H_I(s)\hat{O} H_I(t)  \rho(s)  -H_I(t)\hat{O} H_I(s) \rho(s)\}.
\end{eqnarray}
For $\hat{O} = P_n$ this equation coincides with \eref{eqn:integral_von-Neumann_expand}.
\section*{References}
\bibliographystyle{unsrt}
\bibliography{FPL}
\end{document}